\documentclass[aps,prl,reprint,superscriptaddress,amsmath,amssymb,showpacs,floatfix]{revtex4-1}
\usepackage{graphicx}
\usepackage{color}
\usepackage{CJK}
\usepackage{csquotes}
\usepackage{latexsym}

\begin{document}
\begin{CJK*}{UTF8}{mj}

\title{Time-Delay Induced Dimensional Crossover in the Voter Model}

\author{Mina Kim (김민아)}
\affiliation{Department of Physics, University of Seoul, Seoul 130-743,
Korea}
\author{Jae Dong Noh (노재동)}
\affiliation{Department of Physics, University of Seoul, Seoul 130-743,
Korea}
\affiliation{School of Physics, Korea Institute for Advanced Study,
Seoul 130-722, Korea}
\date{\today}

\begin{abstract}
We investigate the ordering dynamics of the voter model with time-delayed interactions.
The dynamical process in the $d$-dimensional lattice is shown to be equivalent to the first passage problem of a random walker in the $(d+1)$-dimensional strip of a finite width determined by the delay time. 
The equivalence reveals that the time delay leads to the dimensional
crossover from the $(d+1)$-dimensional scaling behavior at a short time to
the $d$-dimensional scaling behavior at a long time. 
The scaling property in both regimes and the crossover time scale are obtained analytically, which are confirmed with the numerical simulation results.
\end{abstract}
\pacs{05.40.-a, 02.50.-r, 02.70.Rr}

\maketitle
\end{CJK*}

Interactions among spatially distributed elements are subjected to a time
delay. Any physical interaction propagates at a finite speed. 
Complex interactions among biological oscillators are mediated by biochemical materials moving at a finite speed~\cite{takamatsu2000time}.  
Such a limited signal propagation speed causes a time delay.
When a delay time is comparable to or larger than characteristic time scales, the time delay brings about rich phenomena. 
It leads to multistable synchronized states for phase
oscillators~\cite{niebur1991collective,kim1997multistability,ernst1998delay,takamatsu2000time,choi2000synchronization,yeung1999time,schuster1989mutual},
amplitude death for coupled limit-cycle
oscillators~\cite{reddy1998time,strogatz1998nonlinear}, stabilization of
periodic orbits in chaotic systems~\cite{pyragas1992continuous}, a resonant
behavior in stochastic
systems~\cite{ohira1995delayed,ohira1999resonance,ohira2000delayed}, dynamic
instability in feedback control systems~\cite{kwon2016information}, a
pattern formation in evolutionary game dynamics~\cite{szolnoki2010dynamically,szolnoki2013decelerated}, and so on.

In this Letter, we investigate the effects of time delay on the ordering dynamics in the context of the voter model~(VM). 
The VM is a prototypical model for opinion dynamics~\cite{liggett2012interacting,glauber1963time,cox1989coalescing,amar1990diffusion,dornic2001critical,krapivsky2010kinetic}. 
In the VM, a voter takes one of the two opinions represented by an Ising spin variable. Each voter selects randomly one of its nearest neighbors and updates its spin state by copying that of the selected neighbor.
Note that the interaction in the VM is {\em instantaneous}.
In a realistic situation, however, an information propagates at a finite speed so that a voter can have an access to the past states of others.
Thus, it is natural to consider the time-delayed interaction in the opinion dynamics, which has never been studied before.
We introduce a time-delayed voter model~(DVM) by incorporating the time delay into the VM and investigate the dynamic scaling behavior. 
We find that the time delay leads to the dimensional crossover: the DVM in
$d$ dimension displays the $(d+1)$-dimensional scaling behavior at a short
time and then $d$-dimensional scaling behavior at a long time.

The time delay makes the dynamics of the DVM non-Markovian. 
Recently, the non-Markovian generalizations of the VM have been considered
by introducing the latency period, the inertia effect, or non-Poissonian
interevent interval distributions ~\cite{Stark:2008ge,Lambiotte:2009ee,Wang:2014hf,takaguchi2011voter}. 
Non-Markovian dynamics of Ising-like systems with time-delayed interactions has also been studied in Refs.~\cite{Choi:1985wy,Huber:2003iz,Kimizuka:2010ch}.  
Our study reveals a new aspect of time-delayed interactions.

The DVM in the $d$-dimensional hypercube $\mathbb{Z}^d$ is defined as follows.
On each site $i\in \mathbb{Z}^d$ resides a voter whose state is represented by an Ising spin variable $s_i = \pm 1$.
During an infinitesimal time interval $dt$, a voter updates its state by
consulting one of its nearest neighbors selected at random with probability $\mu dt$.  
The rate $\mu$ will be set to unity. 
The DVM differs from the voter model by the time delay of length $\tau$ in adopting the opinion of neighbors. 
That is, upon updating at time $t$, a voter at $i$ adopts the opinion of a voter at $j \in \mathcal{N}_i$ at time $(t-\tau)$ where $\mathcal{N}_i$ denotes the set of sites adjacent to $i$.
The delay time $\tau$ is drawn randomly and independently from a
distribution $p_d(\tau)$. In this Letter, we focus on the fixed-$\tau$ distribution with $p_d(\tau) = \delta(\tau- \tau_0)$.
We adopt the random initial condition that $s_i(t)=+1$ or $-1$ with the equal probability for all $i$ and $t\leq 0$.
Because of the time delay, one needs to specify the spin states at negative $t$. 

The ordering dynamics toward the consensus state with all spins up or down is characterized by the spin-spin correlation functions $C_{i,j}(t) = \langle s_i(t)s_j(t)\rangle$ and the domain wall density $\rho(t) = (1-C_{i,j\in \mathcal{N}_i}(t))/2$. Without time delay, the domain wall density decays as
\begin{equation}
\label{VM_domain}
\rho(t) \sim 
\begin{cases}
{t^{-1/2}}, & d=1, \\
{(\ln t)^{-1}}, & d=2,  \\
\end{cases}
\end{equation}
and converges to a finite value for $d> 2$~\cite{krapivsky2010kinetic}. 
These scaling behaviors will be compared with those of the DVM.

\begin{figure}[t]
\includegraphics*[width=\columnwidth]{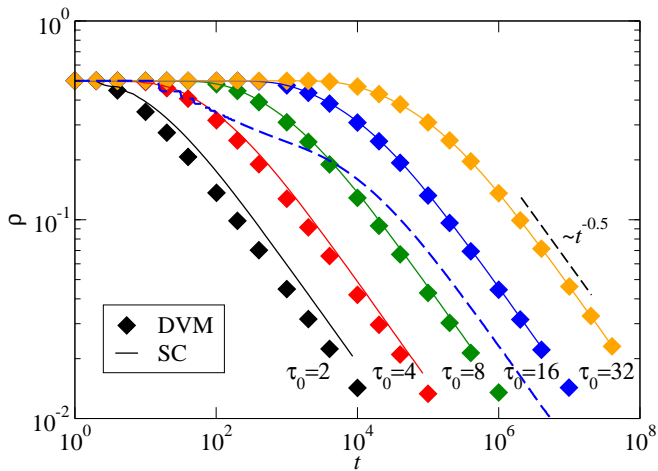}
\caption{\label{fig1} Domain wall density $\rho$ for the DVM with 
fixed-$\tau$ distribution with $\tau_0=2,\cdots,32$. 
The symbols are obtained from the discrete time Monte Carlo simulations 
while the solid lines are from the continuous time SC simulations. Also
shown with the dashed line is the Monte Carlo data for 
$\rho_2$ at $\tau_0=16$.}
\end{figure}

The DVM dynamics on a lattice with $L$ sites can be realized easily in discrete-time Monte Carlo simulations.
One selects a site $i$ and its nearest neighbor site $j$ at random and sets $s_i(t)$ to $s_j(t-\tau)$. 
The time is incremented by $1$ after $L$ updates.
In Fig.~\ref{fig1}, we present the numerical data of $\rho(t)$ for the DVM
in the one-dimensional lattice of $L=2^{11}$ sites under the periodic boundary condition.
The numerical data display an interesting temporal crossover. 
It remains at the initial value $1/2$ at short time, then decays algebraically $\rho(t) \sim t^{-1/2}$ eventually. 
The crossover time increases rapidly as $\tau_0$ increases. 
We also measured the quantity $\rho_2(t) = (1-C_{i,i+2}(t))/2$. 
It also displays the temporal crossover behavior. 
However, its temporal behavior is different from that of $\rho$ qualitatively. 
We will elaborate on the origin for these time-delay induced phenomena.

\begin{figure*}[t]
\includegraphics*[scale=0.21]{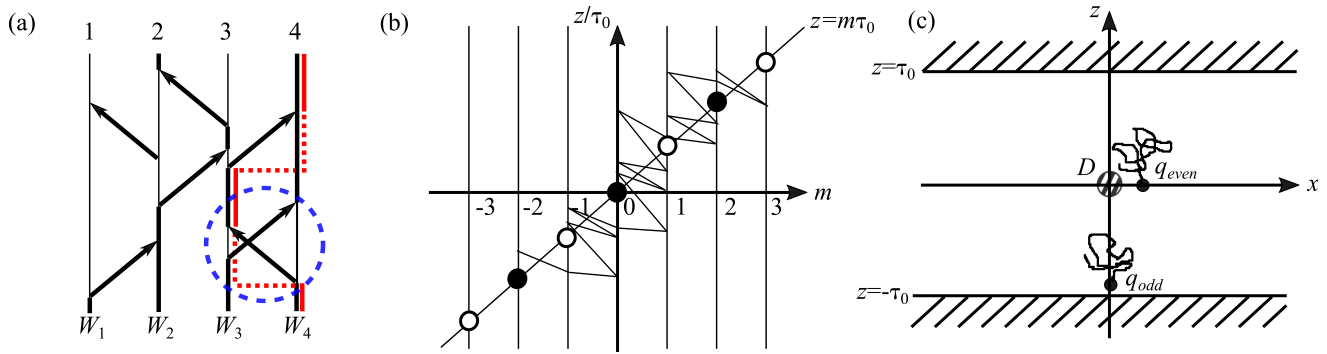}
\caption{\label{fig2} Diagrams for the DVM in one dimension.
(a) Space-time diagram representation of the DVM dynamics. 
The active and inactive periods of $W_4$ are also drawn.
(b) Typical motion of $z$ and $m$ variables defined in the text. 
The filled~(open) circles represent the positions where 
two walkers $W_i$ and $W_j$ with even~(odd) 
$x(0)=x_i(0)-x_j(0)$ can coagulate.
(c) Diffusion of the $R$ walker in the two dimensional $q=(x,z)$ plane of
strip geometry with the target $D$.}
\end{figure*}

The VM without time delay is exactly solvable through the mapping to the coagulating random walks~\cite{cox1989coalescing}. 
We can map the DVM in $d$ dimension to a variant of the coagulating random walks.
The mapping is understood with the help of the $(d+1)$-dimensional space-time diagram~(see Fig.~\ref{fig2}~(a)).
Initially the diagram is composed of vertical lines, each of which corresponds to a timeline of a lattice site in $\mathbb{Z}^d$.
Time runs in the downward direction.
Every time $s_i(t)$ is updated by copying $s_j(t-\tau)$, one adds an arrow from $(i,t)$ to $(j,t-\tau)$. 
The resulting space-time diagram resembles a multilegged ladder with tilted directed rungs.
Given an initial spin configuration at $t\leq 0$ and a space-time diagram, $s_i(t)$ is uniquely determined at all $i$ and $t>0$.
Imagine a walker $W_i$, which starts at $(i,t)$ and climbs up the ladder in the upward~(time-reversed) direction.
Encountering an outgoing rung, the walker hops to the space-time point following the rung. 
Then, one finds that $s_i(t) = s_{x_i(u)}(t-u)$ where $x_i(u)$ denotes the position of the walker $W_i$ after time $u$ with $x_i(0)=i$. 

The $W$ walkers perform a {\em spatial} jump to a neighboring site and a {\em temporal} jump of length $\tau$ at the unit rate in the time-reversed direction. 
When two walkers $W_i$ and $W_j$ meet at the same space-time point, they coagulate into a single walker.
Consequently, the spin-spin correlation function $C_{i,j}(t)$ of the DVM is determined 
by the survival probability $P_{suv}(i,j;t)$ that the two walkers $W_i$ and $W_j$ do not coagulate until time $t$ through the relation
\begin{equation}\label{C_psuv}
C_{i,j}(t) = 1-P_{suv}(i,j;t) \ .
\end{equation}
The domain wall density $\rho(t) = (1-C_{i,j\in \mathcal{N}_i}(t))/2$ is given by 
\begin{equation}\label{rho_psuv}
\rho(t) = \frac{1}{2} P_{suv}(i,j\in \mathcal{N}_i;t)  \ .
\end{equation}

The space-time diagram suggests that the DVM has a slower dynamics than the VM. 
First of all, the diffusion constant of the $W$ walkers is reduced by the
factor $(1+\tau_0)$ because it takes time $(1+\tau_0)$, on average, for each walker to hop. More importantly, the $W$ walkers can cross each other via spatiotemporal jumps as highlighted with a dashed circle in Fig.~\ref{fig2} (a). 
The crossing hinders coagulation and slows down the ordering dynamics. 
In order to stress the role of crossing, we introduce a slightly different but equivalent random walk dynamics.
The walkers are either in an active or inactive state. 
An active walker jumps to a neighboring site at the unit rate, turns into
the inactive state for a time interval $\tau$ drawn from the distribution
$p_d(\tau)$, and becomes active again~(see Fig.~\ref{fig1} (a)).
The coagulation occurs only when two walkers are at the same site and in the active state simultaneously. 
We will call this process the sleeping coagulating random walk process~(SC). 

The domain wall density obtained from time-discretized Monte Carlo simulations of the DVM and the continuous time simulations of the SC using \eqref{rho_psuv} are compared in Fig.~\ref{fig1}.
The two numerical results are in agreement with each other for large $\tau_0$. 
The discrepancy at small values of $\tau_0$ is attributed to the time discretization in the former simulations.

We develop an analytic theory explaining the origin for the crossover behavior.  
For the sake of notational simplicity, the theory is presented for the one dimensional DVM. 
Generalization to higher dimensions is straightforward. 
Consider the motion of two walkers $W_i$ and $W_j$ that are in the active
state at sites $i$ and $j$ initially. 
Their positions are denoted by $x_i$ and $x_j$. 
We adopt the event-driven dynamics.
A unit event consists of the successive periods of being active~(taking time $r$ drawn from the exponential distribution of unit mean) and of being inactive~(taking time $\tau_0$).
Let $n_i$ and $n_j$ denote the numbers of events generated by $W_i$ and $W_j$, respectively. The total number of events is given by $n=n_i+n_j$.
Each walker is assigned to a time label $t_i(n)$ and $t_j(n)$ recording the moment at which the most recent event of each walker is completed. 
They are written as 
\begin{equation}
\label{t_1,2}
\begin{split}
{t_{i}(n)=n_{i}\tau_0 + T_i(n_i)} \\ 
{t_{j}(n)=n_{j}\tau_0 + T_j(n_j)}
\end{split}
\end{equation}
where $T_{i,j}(k)$ denotes the active time period of $W_{i,j}$ which is given by the sum of $k$ independent random variables.
The time evolution is governed by the following rule.
If $t_i(n) <t_j(n)$, the walker $W_i$ generates a new event: $x_i\to x_i\pm 1$, $n_i \to n_i +1$, and ${T}_i\to {T}_i + r$ with a random number $r$ drawn from the exponential distribution of unit mean. 
If $t_j(n) < t_i(n)$, $W_j$ generates a new event similarly.

The dynamics is understood better with the variables $z(n) \equiv {T}_j - {T}_i \in \mathbb{R}$, $m(n) \equiv n_i - n_j \in \mathbb{Z}$, and $x(n) \equiv x_i - x_j \in \mathbb{Z}^{d=1}$. 
If $z(n) > m(n) \tau_0$, then $(z,m,x) \to (z-r,m+1,x\pm 1)$.
Otherwise, $(z,m,x)\to (z+r,m-1,x\pm 1)$.
The variables $z$ and $m$ diffuse along the straight line $z=\tau_0 m$~(see Fig.~\ref{fig2}~(b)). 
Thus, it suffices to consider the variable $z$ only. 
The variable $x$ diffuses independently. 
The process terminates when $x(n) = 0$ and the time labels overlap in the active state. 
We relax the latter condition to $|z(n) - \tau_0 m(n)|<l_0$ with a constant $l_0$ without modifying the long time scaling behavior. 
Note that $m(n)$ and $x(n)$ are incremented by $\pm 1$ at each event so that the parity of $m(n) - x(n)$ is invariant.
It leads to the constraint for the possible values of $m$ at the moment of coagulation: $m$ should be odd~(even) if $x(0)=i-j$ is odd~(even). 
To sum up, the coagulation of two $W$ walkers in the SC is equivalent to the first passage problem of a random walker, denoted as $R$, in the two-dimensional $q=(x,z)$ plane with the targets at $x=0$ and $z=\tau_0 m$ where $m=\pm 1,\pm 3,\cdots$ for odd $x(0)$ and $m=0,\pm 2, \cdots$ for even $x(0)$. 

The system is periodic in the $z$ direction with period $2\tau_0$. Thus, it suffices to consider the semi-infinite strip of width $2\tau_0$ with the periodic~(or reflecting) boundary condition in the $z$ direction~(see Fig.~\ref{fig2}~(c)). 
Using the periodicity, we can place the target at the origin $q=(0,0)$.
In the continuum-$x$ description, the target is replaced by a region $D$ of linear size $l_0$ at the origin $q =(0,0)$.
Then, the survival probability $P_{suv}(i,j;t)$ in the SC is equal to the survival probability $S(q_0,n)$ that the walker $R$ has not visited the target $D$ until $n = t/(1+\tau_0) \simeq t/\tau_0$ steps. 
When $j=i+1$, the walker ${R}$ should start at $q_0 = q_{odd} \equiv (l_0,\tau_0)$.
When $j=i+2$, it should start at $q_0 = q_{even} \equiv (2l_0,0)$.

The strip geometry leads to the dimensional crossover from the two-dimensional behavior for $n \ll \tau_0^2$ to the one-dimensional behavior for $n\gg \tau_0^2$. 
The initial position of the walker $R$ depends on the parity of $i-j$.
Consequently, the correlation function $C_{i,j}(t)$ should have an interesting spatial structure as will be shown below.

The survival probability $S_R(q_0,n)$ of the walker $R$ can be found from the first passage theory~\cite{redner2001guide}.
We will take the continuum approximation so that $q=(x,z)\in \mathbb{R}^2$ and $n\in \mathbb{R}$. 
Let $P_D(q_0,n)$ be the probability that the walker starting at $q_0$ is inside the domain $D$ at time $n$ and $F_D(q_0,n)$ be the probability that it visits the domain $D$ for the first time at $n$. 
Their Laplace transforms $\tilde{P}_D(q_0,s) = \int_0^\infty dn P_D(q_0,n)e^{-sn}$ and $\tilde{F}_D(q_0,s) = \int_0^\infty dn F_D(q_0,n)e^{-sn}$ are related as
\begin{equation}\label{Ftilde}
\tilde{F}_D(q_0,s) = \frac{ \tilde{P}_D (q_0,s) }{\tilde{P}_D(0,s)} .
\end{equation}
Inverting the Laplace transform, one obtains the survival probability from
\begin{equation}
S_{R}(q_0,n) = 1 - \int_0^n F_D(q_0,n')dn' \ .
\end{equation}
In the long time limit, $\int_0^n F_D(q,n')dn'$ can be approximated as $\int_0^\infty F_D(q,n')e^{-n'/n}dn'$~\cite{redner2001guide}, which yields
\begin{equation}\label{S_app}
S_R(q_0,n) \simeq 1- \tilde{F}_D(q_0,s=1/n) \ .
\end{equation}

The random walk problem in the strip geometry can be understood by solving the diffusion equation with the appropriate boundary condition~\cite{Hughes:1996vc,redner2001guide}. 
Instead of doing so, we resort to an intuitive argument in presenting our results. 
The probability $P_D(0,n)$ behaves as
\begin{equation}\label{p0}
P_D(0,n) \sim
\begin{cases}
1 &,~~  n \ll l_0^2 \\
\frac{1}{n} &,~~ l_0^2 \ll n \ll \tau_0^2 \\
\frac{1}{\tau_0 \sqrt{n}}  &,~~ \tau_0^2 \ll n \ .
\end{cases}
\end{equation}
In the second regime, the walker $R$ is not affected by the strip boundary yet and $P_D(0,n)$ displays the two-dimensional scaling. 
When $n\gg \tau_0^2$, the strip becomes equivalent to the one-dimensional 
superlattice of $(2\tau_0) \times (2\tau_0)$ blocks. The walker $R$ jumps between blocks with a scaled diffusion constant $\sim 1/{\tau_0^2}$.
So, $P_D(0,n)$ is given by the product of the probability that ${R}$ is in the central block~($\sim 1/\sqrt{n/\tau_0^2}$) and the fraction of the size of the domain $D$ to the size of the central block~($\sim 1/\tau_0^2$). 
Similarly, one finds that $P_D(q_{odd},n) \sim e^{-a_0 \tau_0^2/n} /n$ for $n \ll \tau_0^2$ and $P_D(q_{odd},n) \sim 1/(\tau_0 \sqrt{n})$ for $n\gg \tau_0^2$. Hereafter, $a_\alpha~(\alpha=0,1,\cdots)$ denotes a nonuniversal constant. 

\begin{figure}[t]
\includegraphics*[width=\columnwidth]{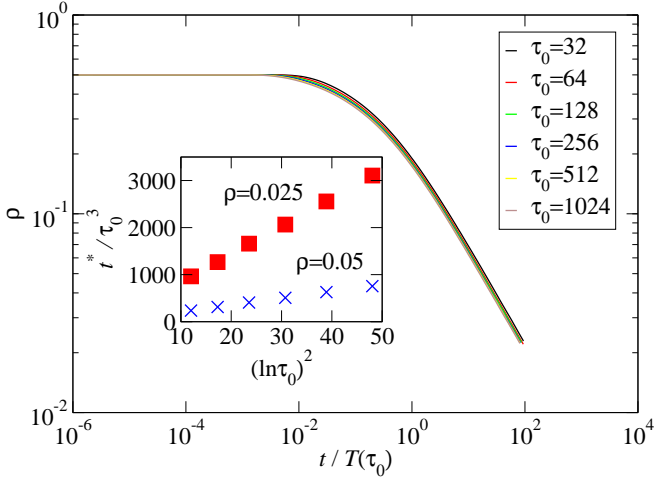}
\caption{\label{fig3} Scaling plot of $\rho(t)$ obtained from the SC
simulations against $t/ [\tau_0^3 (\ln\tau_0)^2]$. 
The inset shows that $t^*/\tau_0^3$, where 
$\rho(t^*)=0.05~(\times)$ and $0.025$~(square), is linear in $(\ln
\tau_0)^2$, which confirms the scaling 
$T(\tau_0) = (1+\tau_0)^3[\ln(1+\tau_0)]^2$.}\end{figure}

Using $P_D(0,n)$ and $P_D(q_{odd},n)$ in \eqref{Ftilde} and \eqref{S_app}, we find that $S_R(q_{odd},n) \simeq 1$ for $n \ll \tau_0^2$ and $S_R(q_{odd},n) \simeq (1+a_1 \sqrt{n}/(\tau_0\ln\tau_0))^{-1}$ for $n\gg \tau_0^2$.
Therefore, the domain wall density $\rho(t) = S_R(q_{odd},t/\tau_0)$ scales as
\begin{equation}\label{rho_scaling}
\rho(t) \simeq \begin{cases}
\frac{1}{2} &,~ t \ll \tau_0^3 \\
\left[1+ \frac{a_1 t}{\tau_0^3(\ln\tau_0)^2}\right]^{-1/2} &,~ t\gg \tau_0^3 .
\end{cases}
\end{equation}
The result explains the crossover behavior observed numerically.
The crossover starts at $t\sim \tau_0^3$. Then, after the intermediate time interval $\tau_0^3 \ll t \ll \tau_0^3 (\ln \tau_0)^2$, the domain wall displays the asymptotic scaling behavior $\rho(t) \sim [t/T(\tau_0)]^{-1/2}$ with $T(\tau_0) = \tau_0^3(\ln\tau_0)^2$. 
We perform the numerical simulations of the SC at large values of $\tau_0$ and confirm the scaling behavior of $T(\tau_0)$ with respect to $\tau_0$ as shown in Fig.~\ref{fig3}.

\begin{figure}[t]
\includegraphics*[width=\columnwidth]{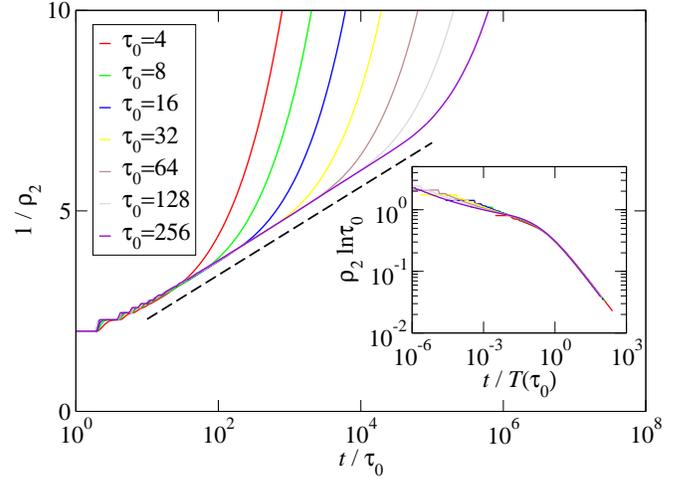}
\caption{\label{fig4} Log-linear plots of $\rho_2(t)$ against $t/\tau_0$ 
for several values of $\tau_{0}$ obtained from the SC simulations. 
The linear segment confirms the logarithmic scaling in the short time
region. The dashed straight line is a guide to an eye. 
The scaling plot in the inset confirms the scaling form in 
\eqref{rho2} in the $t\gg \tau_0^3$ limit.}
\end{figure}

The dimensional crossover is more evident for $\rho_2(t) \equiv [1-C_{i,i+2}(t)]/2 = P_{suv}(i,i+2,t)/2 \simeq S_R(q_{even},n=t/\tau_0)$ with $q_{even}=(2l_0,0)$. 
Note that $P_D(q_{even},n)$ is proportional to $e^{-a_2 l_0^2/n}/n$ for $n \ll l_0^2$, to ${1}/{n}$ for $l_0^2 \ll n \ll \tau_0^2$, and  to ${1}/(\tau_0 \sqrt{n})$ for $n \gg \tau_0^2 $.
Thus, using the Laplace transforms,  we obtain that 
\begin{equation}\label{rho2}
\rho_2(t) \sim \begin{cases}
\left[\ln (t/\tau_0)\right]^{-1} &,~ \tau_0 \ll t \ll \tau_0^3 \\
\left[ \ln \tau_0 + a_3 \sqrt{t/\tau_0^3}\right]^{-1} &,~  t \gg \tau_0^3 .\
\end{cases}
\end{equation}
The two-dimensional logarithmic scaling behavior appears in the regime $\tau_0 \ll t \ll \tau_0^3$~\cite{krapivsky2010kinetic}. 
After the crossover regime at $\tau_0^3 \ll t \ll T(\tau_0)$, it is followed by the one-dimensional scaling $\rho_2(t) \sim (t/\tau_0^3)^{-1/2}$ holds for $t\gg T(\tau_0)$~(See Fig.~\ref{fig4}).

We found that $\rho_2(t)$ decays faster than $\rho(t)$, which implies that a spin becomes more correlated with second nearest neighbors than with nearest neighbors. 
This seemingly counterintuitive result can be understood by considering the space-time diagram representation.
Two voters at sites $i$ and $i+2$ are more likely to have the same opinion by adopting the opinion of the common neighbor at site $i+1$. 
Because of the crossing mechanism, nearest neighbors have less chance. 
The mapping to the $R$ walker problem suggests that all the correlation functions $C_{i,j}$ behave similarly as $C_{i,i+1}$ for $j-i=1~(\mbox{mod2})$ or $C_{i,i+2}$ for $j-i=0~(\mbox{mod 2})$.

We have shown that the time delay in the voter model leads to the dimensional crossover by developing the mapping of the DVM to the problem of the $R$ walker. 
The mapping is valid in any space dimension $d$. 
The $d$-dimensional DVM is dual to the SC, the survival probability of which is obtained from the first passage probability of the $R$ walker in the $(d+1)$-dimensional strip of width $2\tau_0$ with the target at the origin. 
Hence, the dimensional crossover should be present from the $(d+1)$-dimensional behavior for $t\ll \tau_0^3$ to the $d$-dimensional behavior for $t \gg \tau_0^3$. 
In the latter regime, the $(d+1)$-dimensional strip is regarded as the $d$-dimensional lattice of $(2\tau_0)^{d+1}$ blocks.
The spin-spin correlation function $C_{ij}$ displays the spatial pattern, namely the dependence on the parity of the chemical distance between $i$ and $j$, in any lattices which do not include odd loops. 
Detailed studies in higher dimensional lattices will be published elsewhere~\cite{unpub}.
Our study reveals the dimensional crossover as the novel phenomenon induced by the time-delayed interaction. 
It would be interesting to investigate the effect of the time delay on the ordering dynamics of the broader class of models.

\begin{acknowledgments}
This work was supported by the the National Research Foundation of
Korea~(NRF) Grant funded by the Korea government~(MSIP) 
(No.~2016R1A2B2013972).
\end{acknowledgments}

\appendix

\bibliographystyle{apsrev}
\bibliography{paper}
\end{document}